\documentclass[aps,showpacs,nofootinbib]{revtex4}
\usepackage{amsmath} 
\usepackage{amssymb}
\usepackage{graphicx} 
\usepackage{color} 
\usepackage{mathtext}
\usepackage{epstopdf}

\begin{document}
\newcommand{\blue}[1]{\textcolor{blue}{#1}}
\newcommand{\new}{\blue}
\newcommand{\green}[1]{\textcolor{green}{#1}}
\newcommand{\modif}{\green}
\newcommand{\red}[1]{\textcolor{red}{#1}}
\newcommand{\attention}{\red}


\title{Bose--Einstein condensation and muon production in ultra-high energy cosmic ray particle collisions}

\author{V. A. Okorokov} \email{VAOkorokov@mephi.ru; Okorokov@bnl.gov}
\affiliation{National Research Nuclear University MEPhI (Moscow
Engineering Physics Institute), Kashirskoe highway 31, 115409
Moscow, Russia}

\date{\today}

\begin{abstract}
\noindent {\bf Abstract}---Collisions of cosmic ray particles with
ultra-high initial energies with nuclei in the atmosphere open a
wide room for appearing of the novel dynamical features for
multiparticle production processes. In particular, the pion-lasing
behavior driven by Bose--Einstein condensation would result in the
shift to larger multiplicities and, as consequence, could provide,
in general, the enhanced yield of cosmic muons. In the present
work the critical value of the space charged particle density for
onset of Bose--Einstein condensation of the boson (pion)
wave-packets into the same wave-packet state is estimated within
the model with complete multiparticle symmetrization for the
energy domain corresponded to the ultra-high energy cosmic rays
(UHECR). Energy dependence of mean density of charged pions is
evaluated for the cases of absent of the Bose--Einstein effects
and for presence of laser-like behavior of pions. The possible
influence of the Bose--Einstein condensation is discussed for the
muon production in UHECR particle collisions with the atmosphere.
\end{abstract}


\maketitle

\section{Introduction}\label{sec:1}

Measurements of interactions of ultra-high energy cosmic rays
(UHECR), i.e. cosmic ray particles with initial laboratory
energies larger than $10^{17}-10^{18}$ eV, with nuclei in the
atmosphere allow the new unique possibilities for study of
multiparticle production processes at energies (well) above not
only the Large Hadron Collider (LHC) range but future collider on
Earth as well \cite{Okorokov-PAN-82-838-2019}. Possible creation
of a strongly interacting matter under extreme conditions called
also quark-gluon plasma (QGP) can be noted among the important
features of strong interactions of UHECR particles with nuclei in
atmosphere. Due to the air composition and main components of the
UHECR the passage of UHECR particles through atmosphere can be
considered as collision mostly small systems and possibility of
the creation of QGP in such collisions was quantitatively
justified elsewhere \cite{Okorokov-PAN-82-838-2019}.

Muon puzzle is a well-known problem in the physics of high-energy
cosmic rays one of the aspects of which is the muon bundle excess
compared to simulations within available phenomenological models
\cite{NIMA-742-228-2014,EPJWC-210-02004-2019}. The dominant
mechanism for the production of muons in air showers is via the
decay of light charged mesons \cite{PRD-107-094031-2023}. Hadronic
interaction models, continuously informed by new accelerator data,
play a key role in our understanding of the physics driving the
production of extended air showers (EASs) induced by UHECRs in the
atmosphere \cite{arXiv-2205.05845-astro-ph.HE-2022}. Therefore the
study of possible novel features in dynamics of multiparticle
production processes allows the better understanding of the muon
yield in UHECR particle collisions with atmosphere and can shed a
new light on the muon problem.

System with arbitrary number of bosons can undergo a
Bose--Einstein condensation (BEC) due to statistical properties of
quantum system and symmetry of the wave function of a boson state.
In particular, in condensed matter physics, a Bose--Einstein
condensate is a state of matter under some conditions, in which
large fraction of bosons occupy the lowest quantum state (see, for
instance, \cite{book-SM-1-1987}). In that state microscopic
quantum mechanical phenomena, particularly wave function (WF)
interference, become apparent macroscopically, and Bose--Einstein
condensate is described by the WF which is coherent over all
volume of the system under study. Bose--Einstein condensate is
responsible for laser emission, superfluidity and
superconductivity.

As the physical base of BEC is the symmetry of WF
\cite{FE-1-219-1988}, BEC can occur in multi-boson systems at high
particle densities ($n$) and at finite, in the sense of quantum
field theory (QFT), temperatures ($T$), i.e. at $T \sim 0.1-0.2$
GeV \cite{Universe-9-411-2023}. Important feature of BEC at high
$n$ and $T$ is the presence of the finite fraction of particles
which are not in the condensed state
\cite{FE-1-219-1988,Universe-9-411-2023}. That differs BEC at high
$n$ and $T$ from the situation at low $T$ ($T \to 0$) observed,
for instance, in the experiments with ultra cold atoms
\cite{RMP-71-463-1999}. One can note, usually, just the last case
is considered in the text-books as the standard example of BEC,
perhaps, due to its most-known macroscopic manifestations --
superfluidity and superconductivity. Therefore, system with
arbitrary number of bosons can undergo a BEC (a) either by
cooling, (b) or increasing the number density of bosons, (c) or by
increasing the overlap of the multi-boson wave-packet states,
achieved by changing the size of the single-particle wave-packets.

In the real experiments in both the condensed matter physics and
the physics of fundamental interactions the number of particles in
the system is much smaller, on many orders of magnitude, than
macroscopic numbers of particles for which thermodynamic limit is
valid. For example, the number of particles in the state of
Bose--Einstein condensate is $N_{0} \lesssim 10^{7}$ in the modern
experiments with ultra cold atoms \cite{RMP-71-463-1999}.
Therefore, strictly speaking, BEC in the real systems can not be
considered as the phase transition. However, the studies show that
lowest-order correction on finite size of system decreases as
$N^{-1/3}$ with growth of the total number of particles in the
system ($N$) and result for $N_{0}/N$ is indistinguishable from
the exact one obtained by summing explicitly over the excited
states of the harmonic oscillator Hamiltonian already at
$N=10^{3}$ \cite{RMP-71-463-1999}. It was found with help of the
numerical calculations that finite-size effects are significant
only for rather small values $N \lesssim 10^{4}$
\cite{PRA-54-656-1996}. The range of numbers $N \simeq
10^{3}-10^{4}$ corresponds to the total multiplicities of
secondary charged particles in nucleus--nucleus collisions at high
energies.

Regarding the multiparticle production process the mechanisms (b)
and (c) lead to condensation of bosons into the same quantum state
and bosonic, in particular, pion, laser could be created
\cite{PLB-301-159-1993,PRL-80-916-1998,HIP-9-241-1999}. The
resulting average multiplicity of bosons goes to very large, in
general, infinity value if system of secondary bosons undergoes
the BEC, i.e. symmetrization effect is present
\cite{PLB-301-159-1993}. In particular, a Poissonian emmiter that
creates some number of bosons (pions), when symmetrization is
ignored, would emit much larger number of bosons (pions) once
symmetrization was included \cite{PLB-301-159-1993}. Due to decay
mode $\pi^{+} \to \mu^{+}\nu_{\mu}$, $\pi^{-} \to
\mu^{-}\tilde{\nu}_{\mu}$ with $\approx 99.99$\% fraction
\cite{PTEP-2022-083C01-2022} the symmetrization effect for pions
(BEC) could contribute to the enhancement for muon yield obtained
in UHECR particle collisions.

Therefore the study of such novel feature of hadronic processes at
ultra-high energies as BEC seems important for better
understanding of the physics of UHECR and, in particular, it can
be one of the reasons for the excess in muon yield in UHECR
particle interactions with nuclei in atmosphere.

\section{Phenomenology for BEC in nuclear interactions}\label{sec:2}

The basis was described elsewhere \cite{Okorokov-PAN-81-508-2018}
for the using of the standard Mandelstam invariant variable for
nucleon--nucleon / proton-proton pair
$s_{NN/pp}=2m_{N/p}(E_{N/p}+m_{N/p})$ jointly with $E_{N/p}$
within the study of interactions UHECR with nuclei, where
$E_{N/p}$ and $m_{N/p}$ is the energy in the laboratory reference
system and mass of nucleon/proton \cite{PTEP-2022-083C01-2022}.
The relation between $s_{NN}$ and $s_{pp}$ is discussed in details
in the previous work \cite{Okorokov-PAN-82-838-2019}. The energy
range for protons considered in the present paper is
$E_{p}=10^{17}$--$10^{21}$ eV. This range includes the energy
domain corresponded to the Greisen--Zatsepin--Kuzmin (GZK) limit
\cite{Greisen-PRL-16-748-1966} and somewhat expands it because of
the following reasons. On the one hand, both possible
uncertainties of theoretical estimations for the limit values for
UHECR and experimental results, namely, measurements of several
events with $E_{p} > 10^{20}$ eV and the absence of UHECR particle
flux attenuation up to $E_{p} \sim 10^{20.5}$ eV are taking into
account \cite{Okorokov-PAN-81-508-2018}. On the other hand, the
energies corresponding to the nominal value
$\sqrt{\smash[b]{s_{\scriptsize{pp}}}}=14$ TeV of the commissioned
LHC as well as to the parameters for the main international
projects high energy LHC (HE--LHC) with the nominal value
$\sqrt{\smash[b]{s_{\scriptsize{pp}}}}=27$ TeV and Future Circular
Collider (FCC) with $\sqrt{\smash[b]{s_{\scriptsize{pp}}}}=100$
TeV are also included into the aforementioned range of $E_{p}$.
Therefore the estimations below can be useful for both the UHECR
physics and the collider experiments
\cite{Okorokov-PAN-82-838-2019,Okorokov-PAN-81-508-2018}.

The charged particle density is defined as follows:
\begin{equation}
\displaystyle
n_{\scriptsize{\mbox{ch}}}=N_{\scriptsize{\mbox{ch}}} / V,
\label{eq:2.1}
\end{equation}
where $N_{\scriptsize{\mbox{ch}}}$ is the total charged particle
multiplicity, $V$ -- estimation for the volume of the emission
region of the boson under consideration (pions). The critical
value for $n_{\scriptsize{\mbox{ch}}}$
($n_{\scriptsize{\mbox{ch,c}}}$) can be calculated with help of
(\ref{eq:2.1}) and corresponding transition to the critical total
multiplicity ($N_{\scriptsize{\mbox{ch,c}}}$) in this relation.

It should be stressed the physical quantities in r.h.s.
(\ref{eq:2.1}) are model-dependent and, consequently, estimations
for both the charged particle density and the critical one
considered below are also model-dependent.

The equation for the critical value of
$N_{\scriptsize{\mbox{ch}}}$ for 3D case was suggested in
\cite{Okorokov-AHEP-2016-5972709-2016} based on the model for 1D
thermal Gaussian distribution \cite{PLB-301-159-1993}. Application
of the equation from \cite{Okorokov-AHEP-2016-5972709-2016} to the
energy domain $E_{p}=10^{17}-10^{21}$ eV shown the possibility of
Bose--Einstein condensation at least for the interactions of UHECR
particles with nuclei \cite{Okorokov-PAN-82-838-2019}.

Within the present work the following equation for
$N_{\scriptsize{\mbox{ch,c}}}$ is suggested for 3D case based on
the generalized of the pion-laser model for the case of
overlapping wave packets with complete $n$-particle symmetrization
\cite{PRL-80-916-1998,HIP-9-241-1999}:
\begin{eqnarray}
\displaystyle
N_{\scriptsize{\mbox{ch,c}}}&=&\frac{1}{\eta}\biggl(\frac{1+X+\sqrt{1+2X}}{2}\biggr)^{3/2}, \\
X& \equiv&
2m_{\pi}T_{\scriptsize{\mbox{eff}}}R_{\scriptsize{\mbox{eff}}}^{2},
~~~~~
T_{\scriptsize{\mbox{eff}}}=T+\frac{\Delta_{p}^{2}}{2m_{\pi}},
~~~~~
R_{\scriptsize{\mbox{eff}}}^{2}=R_{m}^{2}+\frac{1}{2\Delta_{p}^{2}}\frac{T}{T_{\scriptsize{\mbox{eff}}}}.\nonumber
\label{eq:2.2}
\end{eqnarray}
Here $T_{\scriptsize{\mbox{eff}}}$ and
$R_{\scriptsize{\mbox{eff}}}$ is effective temperature and radius
of the emission region (source), $\Delta_{p}$ is the momentum
spread of the emitted bosons (pions) dependent, in general, on
type of collision, $\eta=0.25$ is the fraction of the pions to be
emitted from a static Gaussian source (1-st generation particle)
within unit of rapidity \cite{PLB-301-159-1993}, $R_{m}$ is the
estimation of the source size (radius), $T \approx
T_{\scriptsize{\mbox{ch}}}$ is the source temperature supposed
equal to the parameter value at chemical freeze-out
\cite{Okorokov-AHEP-2016-5972709-2016}.

In accordance with \cite{Okorokov-AHEP-2016-5972709-2016} the same
analytic energy dependence of $T_{\scriptsize{\mbox{ch}}}$
\cite{PRC-73-034905-2006} is suggested for various ($p+p\,/\,A+A$)
strong interaction processes. Based on \cite{PLB-301-159-1993} the
one value $\Delta_{p}=0.250$ GeV was used in our previous works
\cite{Okorokov-PAN-82-838-2019,Okorokov-AHEP-2016-5972709-2016}
for any types of collisions. But some larger $\Delta_{p}$ can be
expected for $A+A$ collisions than that in $p+p$ interactions
because of, in general, influence of strongly interacting
environment. Thus the empirical values $\Delta_{p}=0.250$ and
0.375 GeV are used for $p+p$ and $A+A$ collisions respectively
within the present study. More rigorous choice of the values of
$\Delta_{p}$ in various collisions requires the additional
consideration for reliable justification.

In the present work the following two analytic functions are used
for the parametrization of the energy-dependent average total
charged particle multiplicity in $p+p$ collisions. Hybrid function
\begin{equation}
\displaystyle \langle N_{\scriptsize{\mbox{ch}}}^{pp}\rangle=
1.60-0.03\ln\varepsilon_{pp}+0.18\ln^{2}\varepsilon_{pp}+0.03\,\varepsilon^{0.29}_{pp}
\label{eq:2.3}
\end{equation}
was deduced within participant dissipating energy (PDE) approach
\cite{PRD-93-054046-2016} with $\varepsilon_{NN/pp} \equiv
s_{NN/pp}/s_{0}$, $s_{0}=1$ GeV$^{2}$. Application of the Quantum
Chromodynamics (QCD) as the model independent quantum--field basis
of the theory of strong interactions seems important for the study
of multiparticle production processes at ultra--high energies, in
particular, for a features of UHECR particle collisions. The QCD
inspired analytic function \cite{JPGNNP-37-083001-2010}
\begin{equation}
\displaystyle \langle
N_{\scriptsize{\mbox{ch}}}^{pp}\rangle=\langle
N_{\scriptsize{\mbox{ch,F}}}\rangle+N_{0} \label{eq:2.4}
\end{equation}
is considered within the present work as the second case of
analytic parametrization for $p+p$ for completeness information,
where $N_{0}$ is, in general, the free parameter with chosen value
$N_{0}=2.20 \pm 0.19$ \cite{JPGNNP-37-083001-2010} and $\langle
N_{\scriptsize{\mbox{ch,F}}}\rangle$ is the mean charged particle
multiplicity in quark jet produced in $e^{+}e^{-}$ annihilation.
The perturbative QCD (pQCD) in the
next--to--next--to--next--to--leading order (N$^{3}$LO) allows the
analytic solution \cite{PLB-459-341-1999,PR-349-301-2001} for
$\langle N_{\scriptsize{\mbox{ch,F}}}\rangle$
\begin{eqnarray}
\displaystyle \label{eq:2.5} \langle
N_{\scriptsize{\mbox{ch,F}}}\rangle&=&\frac{K_{\scriptsize{\mbox{LHPD}}}}{r_{0}}Y^{-a_{1}c^{2}}\exp\bigl[2c\sqrt{Y}+\delta_{\scriptsize{\mbox{F}}}(Y)\bigr], \\
\delta_{\scriptsize{\mbox{F}}}(Y)&
=&\frac{c}{\sqrt{Y}}\biggl[r_{1}+2a_{2}c^{2}+\frac{\beta_{1}}{2\beta_{0}^{2}}(\ln2Y
+2)\biggr]+\frac{c^{2}}{Y}\biggl[a_{3}c^{2}+\frac{r_{1}^{2}}{2}+r_{2}-\frac{a_{1}\beta_{1}}{2\beta_{0}^{2}}(\ln2Y
+1)\biggr].\nonumber
\end{eqnarray}
Here $K_{\scriptsize{\mbox{LHPD}}}$ is an overall normalization
constant due to local hadron--parton duality (LHPD), $Y \equiv
\ln(k_{0}\sqrt{s_{pp}}/2\Lambda)$, $c=\sqrt{N_{c}/\pi \beta_{0}}$,
$N_{c}$ is the number of colors, $\beta_{i}$, $i=0,1$ are the
($i+1$)--loop coefficients of the $\beta$-function
\cite{PTEP-2022-083C01-2022}, $k_{0}=0.35 \pm 0.01$
\cite{JPGNNP-37-083001-2010} and the parameters
$K_{\scriptsize{\mbox{LHPD}}}$, $\Lambda$ and $\forall\,i: a_{i},
r_{i}$ depend on number of active quark flavors $N_{f}$. Their
numerical values can be found in
\cite{PR-349-301-2001,EPJC-35-457-2004}.

For the case of $A+A$ collisions it is supposed that the
dependence of total charged particle multiplicity on collision
energy is approximated by the following two analytic functions.
Hybrid function
\begin{equation}
\displaystyle \xi^{-1}\langle
N_{\scriptsize{\mbox{ch}}}^{AA}\rangle=
-0.577+0.394\ln\varepsilon_{NN}+0.213\ln^{2}\varepsilon_{NN}+0.005\,\varepsilon^{0.55}_{NN}
\label{eq:2.6}
\end{equation}
was deduced within PDE approach \cite{PRD-93-054046-2016} as well
as the formula (\ref{eq:2.3}) for $p+p$, where $2\xi=\langle
N_{\scriptsize{\mbox{part}}}\rangle$ and
$N_{\scriptsize{\mbox{part}}}$ number of participants. The
function
\begin{equation}
\displaystyle \xi^{-1}\langle
N_{\scriptsize{\mbox{ch}}}^{AA}\rangle=
1.962+0.512\,\varepsilon^{0.15}_{NN}\ln\varepsilon_{NN}
\label{eq:2.7}
\end{equation}
is the fit to the Relativistic Heavy Ion Collider (RHIC) and ALICE
facility (A Large Ion Collider Experiment) data in wide energy
domain up to the $\sqrt{s_{NN}}=2.76$ TeV \cite{PLB-726-610-2013}.
It is important to note that the function (\ref{eq:2.7})
reasonably describes heavy ion (Au+Au, Pb+Pb) results and
experimental points obtained for Cu+Cu collisions, i.e. for
interactions of moderate nuclei \cite{PLB-726-610-2013}.

The quantitative estimations of $V$ have been derived with help of
the approach from
\cite{Okorokov-PAN-82-838-2019,Okorokov-AHEP-2016-5972709-2016,Okorokov-AHEP-2015-790646-2015},
namely, with help of the extrapolation of the results obtained for
pion femtoscopy on the ultra-high energy domain. As it was
stressed in \cite{Okorokov-AHEP-2016-5972709-2016} the
phenomenology used here allows the estimation of the upper
boundary only for true value of the ratio
$n_{\scriptsize{\mbox{ch}}}/n_{\scriptsize{\mbox{ch,c}}}$ with
additional uncertainty $\sim 3\sqrt{\pi/2}$ due to various
definitions of the emission region volume applied for the
estimation of $n_{\scriptsize{\mbox{ch}}}$ (cylindrical-like shape
of the source) and for $n_{\scriptsize{\mbox{ch,c}}}$
(spherically-symmetrical shape of the source). Detailed discussion
of this issue can be found elsewhere
\cite{Okorokov-AHEP-2016-5972709-2016} and aforementioned feature
should be taken into account for all results in Sec. \ref{sec:3}
and \ref{sec:4}.

The states of system with arbitrary number of bosons $n$, normalized to unity, can be
written as follow:
\begin{equation}
\displaystyle |\alpha_{1}, ..., \alpha_{n}\rangle =
\biggl(\sum\limits_{\tau^{(n)}}\prod\limits_{k=1}^{n}\langle
\alpha_{k}|\alpha_{\tau_{k}}\rangle\biggr)^{1/2}\alpha_{n}^{+}...\alpha_{1}^{+}|0\rangle.\nonumber
\end{equation}
Here $\displaystyle
\alpha_{i}^{+}=\int \frac{d^{3}{\bf p}}{(\pi
\sigma_{i})^{3/4}}\exp\bigl[-({\bf p}-{\bf
p}_{0i})^{2}/2\sigma_{i}^{2}-i{\bf x}_{0i}({\bf p}-{\bf
p}_{0i})+i\omega({\bf p})(t-t_{0i})\bigr]a_{{\bf p}}^{+}$ is the creation operator for
single-particle wave packet in momentum space with the center characterized by the vector ${\bf x}_{0i}$
(${\bf p}_{0i}$) in the coordinate (momentum) space, the width $\sigma_{i}$ in momentum space and the production time $t_{i}$, $a_{{\bf p}}^{+}$ is the creation operator for boson (pion) with momentum ${\bf p}$, $\forall\,i=1-n: \alpha_{i}=({\bf
x}_{0i},{\bf p}_{0i},\sigma_{i},t_{0i})$ corresponds to the single-particle wave packet with given set of parameters, summation is performed over the set $\tau^{(n)}$ of all the permutations of the indexes $\{1, 2, ..., n\}$ of $n$-bosonic state and $\tau_{k}$ denotes the index that replaces the index $k$ in a given permutation from $\tau^{(n)}$ \cite{PRL-80-916-1998}.
Within the generalization of the pion laser model to the case of
overlapping wave packets and assumptions about the increase of the
probability of boson emission in the presence of other source in the vicinity
(analogue of induced emission) the following equation is obtained
for the density matrix of the $n$-particle bosonic state \cite{PRL-80-916-1998,HIP-9-241-1999}:
\begin{equation}
\displaystyle
\rho_{n}(\alpha_{1},...,\alpha_{n})=\frac{1}{\mathcal{N}(n)}\prod\limits_{i=1}^{n}
\rho_{1}(\alpha_{i})\biggl(\sum\limits_{\tau^{(n)}}\prod\limits_{k=1}^{n}\langle
\alpha_{k}|\alpha_{\tau_{k}}\rangle\biggr), \label{eq:2add.1}
\end{equation}
where the coefficient $\mathcal{N}(n)$ is determined
from the normalization condition of density matrix of the quantum system.

The density matrix
(\ref{eq:2add.1}) corresponds to the hypothesis that the creation of a
boson has a larger probability in a state already filled by
another boson and describes a quantum system of wave packets with
induced emission. The intensity of this induced emission, i.e. the number of emitted bosons is controlled
by the degree of overlap of single-particle wave packets -- the degree of
symmetrization of WF of the $n$-particle state. Excess in $n$-particle density matrix
$\rho_{n}(\alpha_{1},...,\alpha_{n})$ with accounting for the symmetrization effect over
corresponding value for the case of complete absence of overlap of single-particle wave packets, i.e. for completely asymmetrical WF, defined as
\begin{equation}
\displaystyle
\rho_{n}(\alpha_{1},...,\alpha_{n})\frac{\mathcal{N}(n)}{\prod\limits_{i=1}^{n}
\rho_{1}(\alpha_{i})}=\sum\limits_{\tau^{(n)}}\prod\limits_{k=1}^{n}\langle
\alpha_{k}|\alpha_{\tau_{k}}\rangle. \label{eq:2add.2}
\end{equation}
The value of the weight (\ref{eq:2add.2}) varies from 1 for the completely asymmetrical case
(complete absence of overlap of wave packets) up to $n!$ at full
$n$-particle symmetrization \cite{PRL-80-916-1998,HIP-9-241-1999}, i.e. for the emission of all $n$ particles in identical wave states
packets (maximal overlap of all $n$ single-particle wave
packages). Thus, the overlap of the wave packets in
multi-boson states can, under certain conditions, lead to
to BEC and, as a consequence, to a significant increase of the multiplicity of
secondary particles in the case of $n$-particle symmetrization of WF.

The influence of the Bose--Einstein condensation on secondary
boson multiplicity was considered in \cite{HIP-9-241-1999}, in
particular, for the special case of the Poissonian multiplicity
distribution. The Poissonian distribution with mean $n_{0}$ is the
following for the case when the Bose--Einstein effects are
switched off
\begin{equation}
\displaystyle \mathcal{P}_{n}^{(0)}=(n_{0}^{n}/n!)\exp(-n_{0}).
\label{eq:2.8}
\end{equation}
Then the probability distribution for the special case of the rare
Bose gas, i.e. $X \gg 1$, with taking into account the BEC, allows
the analytic formula \cite{HIP-9-241-1999}
\begin{equation}
\displaystyle
\mathcal{P}_{n}=\mathcal{P}_{n}^{(0)}\biggl[1+\frac{n(n-1)-n_{0}^{2}}{2(2X)^{3/2}}\biggr]
\label{eq:2.9}
\end{equation}
with the mean value \cite{HIP-9-241-1999}
\begin{equation}
\displaystyle n=n_{0}\biggl[1+\frac{n_{0}}{(2X)^{3/2}}\biggr].
\label{eq:2.10}
\end{equation}
As seen the relative increase of mean value due to influence of
BEC is $(\delta n)_{\scriptsize{\mbox{BEC}}}=(n-n_{0})/n_{0}
\propto X^{-3/2}$ as the function of the $X$ variable for the
specific aforementioned case.

It can be noted that the relations (\ref{eq:2.9}) and
(\ref{eq:2.10}) were deduced in \cite{HIP-9-241-1999} with help of the
density matrix, which is the most general form of description of quantum systems underlying quantum
statistics. Furthermore any limitations were absent for kinematic parameters (4-momentum) of the initial state particles within the approach from \cite{HIP-9-241-1999}. All these allows the suggestion for correctness of the equations (\ref{eq:2.9}) and (\ref{eq:2.10}) for the domain of ultra-high initial energies under consideration, respectively, for the specific case of Poissonian multiplicity distribution.

\section{Results and discussion}\label{sec:3}

\begin{figure*}
\includegraphics[width=14.0cm,height=14.0cm]{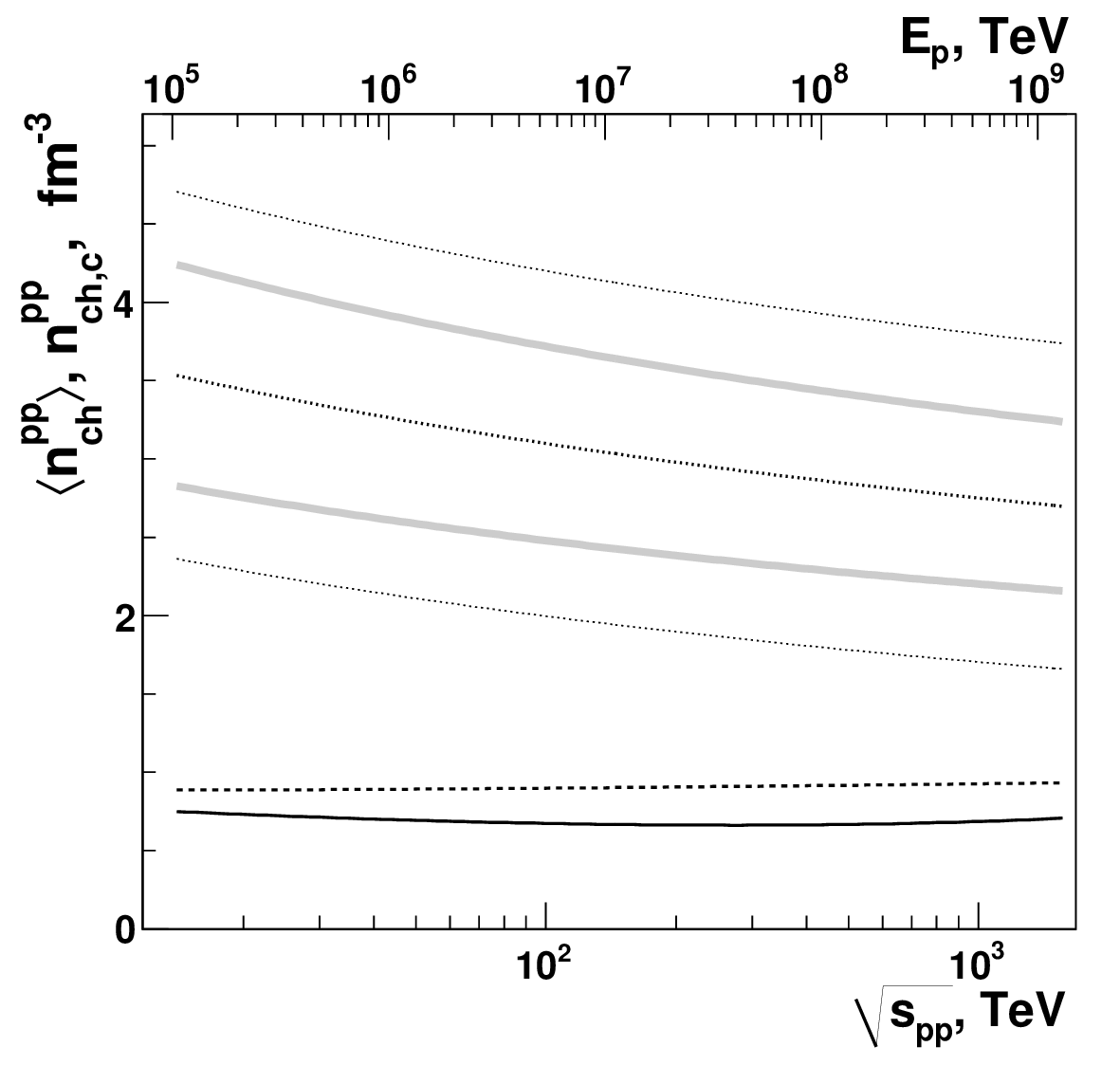}
\vspace*{8pt} \caption{Energy dependence of $\langle
n_{\scriptsize{\mbox{ch}}}\rangle$ and critical parameter within
approach from \cite{PRL-80-916-1998,HIP-9-241-1999} in $p+p$
collisions. Solid line corresponds to the hybrid approximation
(\ref{eq:2.3}) of $\langle N_{\scriptsize{\mbox{ch}}}^{pp}\rangle$
and dashed line is for QCD inspired function (\ref{eq:2.4}) with
N$^{3}$LO pQCD equation (\ref{eq:2.5}). Critical density is shown
by dotted line with its statistical uncertainty levels represented
by thin dotted lines. The heavy grey lines correspond to the
systematic $\pm 1$ s.d. of the quantity calculated by varying of
$\eta$ on $\pm 0.05$.}\label{fig:1}
\end{figure*}

Fig. \ref{fig:1} shows $\langle
n_{\scriptsize{\mbox{ch}}}^{pp}\rangle$ and
$n_{\scriptsize{\mbox{ch,c}}}^{pp}$ depends on energy parameters
in $p+p$ collisions. Here and in future figures two axes are shown
for completeness of the information, namely, the lower axis is the
center-of-mass collision energy for proton-proton pair and the
upper axis is the the energy of incoming particle in the
laboratory reference system. The $\langle
n_{\scriptsize{\mbox{ch}}}^{pp}\rangle(s_{pp})$ is calculated with
help of the hybrid approximation (\ref{eq:2.3}) of $\langle
N_{\scriptsize{\mbox{ch}}}^{pp}\rangle$ (solid line) as well as
the QCD inspired function (\ref{eq:2.4}) with $\langle
N_{\scriptsize{\mbox{ch,F}}}\rangle$ estimated by (\ref{eq:2.5})
up to the N$^{3}$LO pQCD (dashed line). Quantity
$n_{\scriptsize{\mbox{ch,c}}}$ estimated within approach from
\cite{PRL-80-916-1998,HIP-9-241-1999} is shown by dotted line with
its statistical uncertainty levels represented by thin dotted
lines. The heavy grey lines correspond to the systematic $\pm 1$
s.d. of $n_{\scriptsize{\mbox{ch,c}}}$ calculated by varying of
$\eta$ on $\pm 0.05$ in (\ref{eq:2.2}). The
$n_{\scriptsize{\mbox{ch}}}$ in $p+p$ is noticeably smaller than
its critical value at collision energies up to $\sqrt{s_{pp}} \sim
1$ PeV for any analytic approximations for $\langle
N_{\scriptsize{\mbox{ch}}}^{pp}\rangle$ considered here (Fig.
\ref{fig:1}).

The estimations for the parameter $\langle
n_{\scriptsize{\mbox{ch}}}^{pp}\rangle$, available based on
measurements, are presented in
\cite{Okorokov-AHEP-2016-5972709-2016} and are limited to the
range $\sqrt{s_{pp}} \leq 7$ TeV, which is significantly smaller
than the lower limit of the energy range under consideration
$E_{\scriptsize{\mbox{min}}}=10^{17}$ eV $\longleftrightarrow
\sqrt{s_{\scriptsize{\mbox{min}}}} \approx 13.7$ TeV
\cite{Okorokov-PAN-82-838-2019}. The LHC has already collected
data for $p+p$ collisions at $\sqrt{s_{pp}}=13$ and 13.6 TeV,
which is (very) close to $\sqrt{s_{\scriptsize{\mbox{min}}}}$,
however for these $\sqrt{s_{pp}}$ there are no experimental
results yet, in particular, for the geometry of the secondary pion
emission region in three-dimensional case, and the available data
from one-dimensional analyzes are not allow us to estimate the
volume of the source for realistic (cylinder-like) shape.
Therefore, the new results of accelerator experiments for $\langle
N_{\scriptsize{\mbox{ch}}}^{pp}\rangle$ and $V$ in the multi-TeV
energy region are important for checking and future improving of
the phenomenological approach suggested in the present work. The
data acquisition is expected at nominal value $\sqrt{s_{pp}}=14$
TeV of the commissioned LHC. In the longer term, the
implementation of the international projects of high energy mode
LHC (HE--LHC) and Future Circular Collider with proton and ion
beams (FCC--hh) is expected to provide data with high statistical
precision for $p+p$ collisions at $\sqrt{s_{pp}}=27$ TeV
\cite{EPJST-228-1109-2019} and 100 TeV \cite{EPJST-228-755-2019}
respectively.

Current information for the mass composition of UHECR is limited
and characterized by significant errors
\cite{arXiv-2205.05845-astro-ph.HE-2022,PPNP-63-293-2009,PRD-103-103009-2021},
which is mainly due to uncertainties in models of hadron
interactions used to describe extended air showers
\cite{PPNP-63-293-2009}. Despite the fact that in some cases --
determination of the mass composition of primary UHECR based on
measuring of the depth of maximum muon production
($X^{\mu}_{\scriptsize{\mbox{max}}}$) -- certain models, namely,
EPOS--LHC predict the contribution of components with $4 < \langle
\ln\,A\rangle \leq 8$ at $2 \times 10^{19} \leq E_{p} \leq 6
\times 10^{19}$ eV \cite{PRD-90-012012-2014}, the consensus of the
main parts of available experimental and phenomenological data
allows us to conclude that the mass composition of the UHECR is
almost completely defined by components down to the nucleus
${}^{56}\mbox{Fe}^{26+}$ taking into account (large) errors in the
energy range under consideration $E_{p}=10^{17}$ -- $10^{21}$ eV,
i.e. down to the nuclei with $A \lesssim 60$
\cite{arXiv-2205.05845-astro-ph.HE-2022,PPNP-63-293-2009,PRD-103-103009-2021},
where $A$ is the mass number. On the other hand, as it was
stressed at the study of global characteristics of nuclear
collisions at ultra-high energies \cite{Okorokov-PAN-82-838-2019},
the free parameter values in (\ref{eq:2.6}), (\ref{eq:2.7}) for
$A+A$ have been obtained for heavy\footnote{This term is used in
relation to the entire Periodic table of the elements and in the
sense corresponding to modern accelerator physics, i.e. nuclei
with $A \gtrsim 200$ are meant under heavy ones, for example,
${}^{197}\mbox{Au}^{79+}$, ${}^{207}\mbox{Pb}^{82+}$ etc. as noted
in the explanation to (\ref{eq:2.7}).} ion collisions mostly and
usually for the most central bin. Therefore, strictly speaking,
the results obtained within the present work and considered below
are for symmetric ($A+A$) nuclear collisions for heavy and
moderate, down to the ${}^{64}\mbox{Cu}^{29+}$, ions. Its
applicability for light nuclei which are the main components of
UHECR requires the additional justification and careful
verification\footnote{As seen the lightest nucleus taken into
account, for instance, by the analytic function (\ref{eq:2.7}) is
close to the heaviest component of UHECR. This observation can be
considered as some support and positive argument in favour of
applicability of the present study to the UHECR at least on
qualitative level.}.

\begin{figure*}
\includegraphics[width=14.0cm,height=14.0cm]{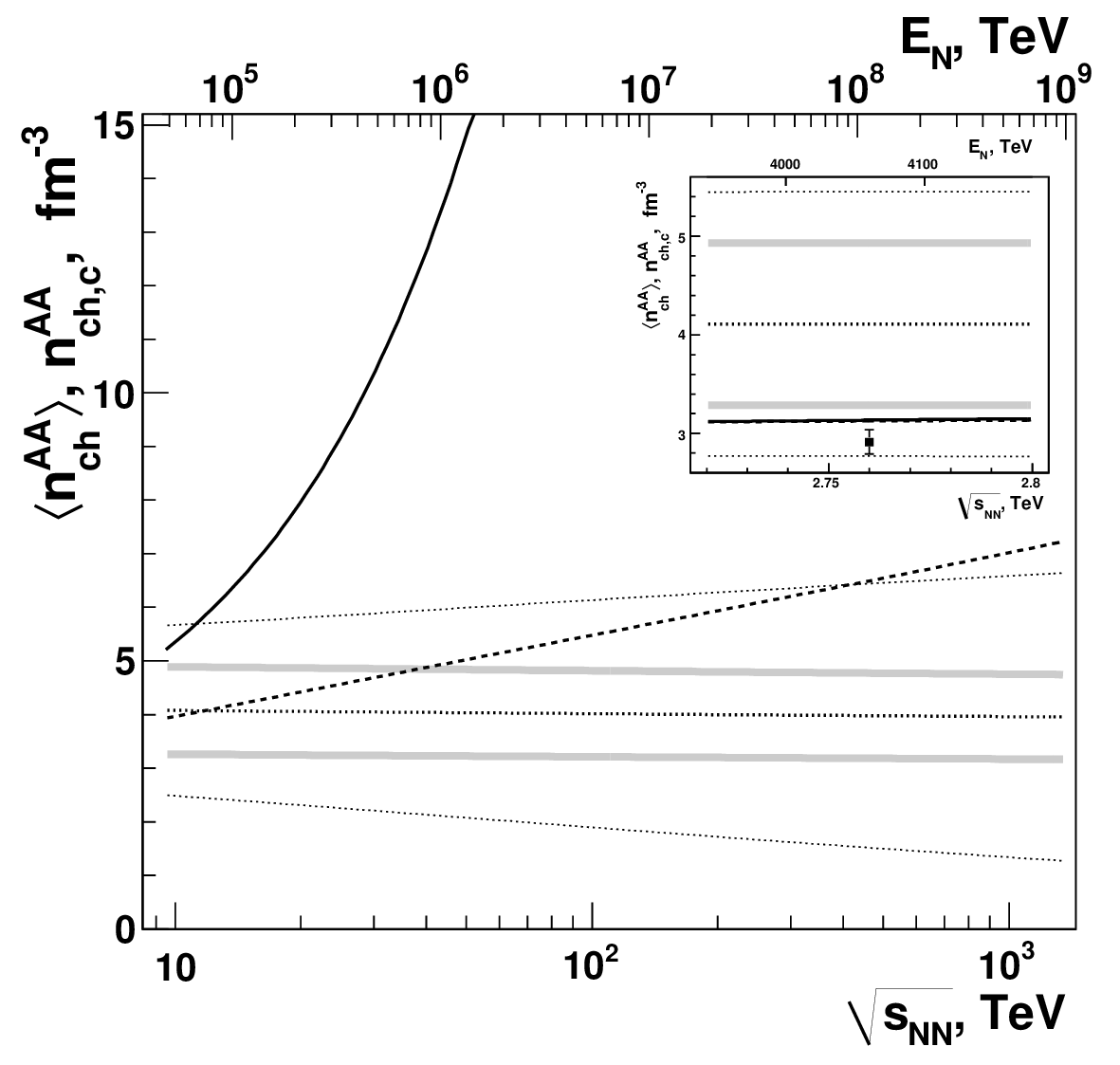}
\vspace*{8pt} \caption{Energy dependence of $\langle
n_{\scriptsize{\mbox{ch}}}\rangle$ and critical parameter within
approach from \cite{PRL-80-916-1998,HIP-9-241-1999} in symmetric
($A+A$) ion collisions. Solid line corresponds to the hybrid
approximation (\ref{eq:2.6}) of $\langle
N_{\scriptsize{\mbox{ch}}}^{AA}\rangle$ and dashed line is for
equation (\ref{eq:2.7}). The notations of the curves for
$n_{\scriptsize{\mbox{ch,c}}}^{AA}$ are identical to that for
$n_{\scriptsize{\mbox{ch,c}}}^{pp}$ in Fig. \ref{fig:1}. The inner
panel: the point is the estimation for $\langle
n_{\scriptsize{\mbox{ch}}}^{AA}\rangle$ at $\sqrt{s_{NN}}=2.76$
TeV \cite{Okorokov-AHEP-2016-5972709-2016} and curves calculated
within the present work for narrow energy range
$\sqrt{s_{NN}}=2.72-2.80$ TeV close to the point.}\label{fig:2}
\end{figure*}

Taking into account this consideration the secondary boson (pion)
densities $\langle n_{\scriptsize{\mbox{ch}}}^{AA}\rangle$ and
$n_{\scriptsize{\mbox{ch,c}}}^{AA}$ are studied for symmetric
($A+A$) nuclear collisions in energy domain corresponded to the
UHECR. In Fig. \ref{fig:2} the parameters $\langle
n_{\scriptsize{\mbox{ch}}}^{AA}\rangle$ and
$n_{\scriptsize{\mbox{ch,c}}}^{AA}$ are shown in dependence on
energy parameters for $A+A$ collisions. Solid line corresponds to
the hybrid approximation (\ref{eq:2.6}) of $\langle
N_{\scriptsize{\mbox{ch}}}^{AA}\rangle$ and dashed line is for
equation (\ref{eq:2.7}). The notations of the curves for
$n_{\scriptsize{\mbox{ch,c}}}^{AA}$ are identical to that for
$n_{\scriptsize{\mbox{ch,c}}}^{pp}$ in Fig. \ref{fig:1}. Here and
below for the case of $A+A$ collisions the value of $\xi$
corresponds to the heavy ion type (${}^{208}\mbox{Pb}^{82+}$) of
incoming particles from \cite{PLB-726-610-2013}. The quantitative
study based on the available measurements for lighter nuclei is in
the progress. The $\langle n_{\scriptsize{\mbox{ch}}}^{AA}\rangle$
is larger than critical value for charged particle density in
$A+A$ collisions at any energies under consideration ($E_{N} \geq
10^{17}$ eV) for both equations (\ref{eq:2.6}) and (\ref{eq:2.7})
used for the approximation of $\langle
N_{\scriptsize{\mbox{ch}}}^{AA}\rangle(s_{NN})$ if only mean curve
for $\langle n_{\scriptsize{\mbox{ch,c}}}^{AA}\rangle$ is taken
into account in Fig. \ref{fig:2}. As the nuclear collisions allows
the laser-like regime for multiparticle production at ultra-high
energies the possible influence of BEC is studied for multipion
final state in the specific case of Poissonian distribution for
energy domain corresponded to UHECR collisions with atmosphere.

The estimations for the parameter $\langle
n_{\scriptsize{\mbox{ch}}}^{AA}\rangle$, available based on
measurements, are presented in
\cite{Okorokov-AHEP-2016-5972709-2016} and are limited to the
range $\sqrt{s_{NN}} \leq 2.76$ TeV, which is significantly
smaller than $\sqrt{s_{\scriptsize{\mbox{min}}}}$. For
completeness of information, Fig. \ref{fig:2} (inner panel) shows
$\langle n_{\scriptsize{\mbox{ch}}}^{AA}\rangle$ at
$\sqrt{s_{NN}}=2.76$ TeV \cite{Okorokov-AHEP-2016-5972709-2016} in
comparison with the curves for
$n_{\scriptsize{\mbox{ch,c}}}^{AA}(s_{NN})$ obtained within the
present work. The estimation of $\langle
n_{\scriptsize{\mbox{ch}}}^{AA}\rangle$ at $\sqrt{s_{NN}}=2.76$
TeV is (very) close to the lower end of the range
$[n_{\scriptsize{\mbox{ch,c}}}^{AA}-\Delta_{\scriptsize{\mbox{stat}}}n_{\scriptsize{\mbox{ch,c}}}^{
AA};
n_{\scriptsize{\mbox{ch,c}}}^{AA}+\Delta_{\scriptsize{\mbox{stat}}}n_{\scriptsize{\mbox{ch,c}}}^{AA}
]$ where the onset of BEC is possible,
$\Delta_{\scriptsize{\mbox{stat}}}n_{\scriptsize{\mbox{ch,c}}}^{AA}$
is statistical error of $n_{\scriptsize{\mbox{ch,c}}}^{AA}$. Thus,
the appearance of BEC seems unlikely even for collisions of heavy
(Pb+Pb) nuclei at $\sqrt{s_{NN}}=2.76$ TeV within the generalized
pion laser model. This conclusion is in good agreement with the
results of searching for signatures of BEC using multipion
correlations in Pb+Pb interactions at $\sqrt{s_{NN}}=2.76$ TeV
\cite{PRC-93-054908-2016}. In difference with $p+p$, obtaining of
estimations for $\langle n_{\scriptsize{\mbox{ch}}}^{AA}\rangle$
in nucleus--nucleus interactions, based on data of accelerator
experiments at the considered ultra-high energies, can only be
expected in the long term when the FCC--hh project for ion
beams\footnote{In this work it does not consider strongly
asymmetric $p$+Pb collisions, for which it is possible to achieve
$\sqrt{s_{NN}}=17$ TeV already within the framework of the HE--LHC
project \cite{EPJST-228-1109-2019}.} will be commissioned, in
which, in particular, it is planned to study Pb+Pb at
$\sqrt{s_{NN}}=39$ TeV \cite{EPJST-228-755-2019}.

\begin{figure*}
\includegraphics[width=14.0cm,height=14.0cm]{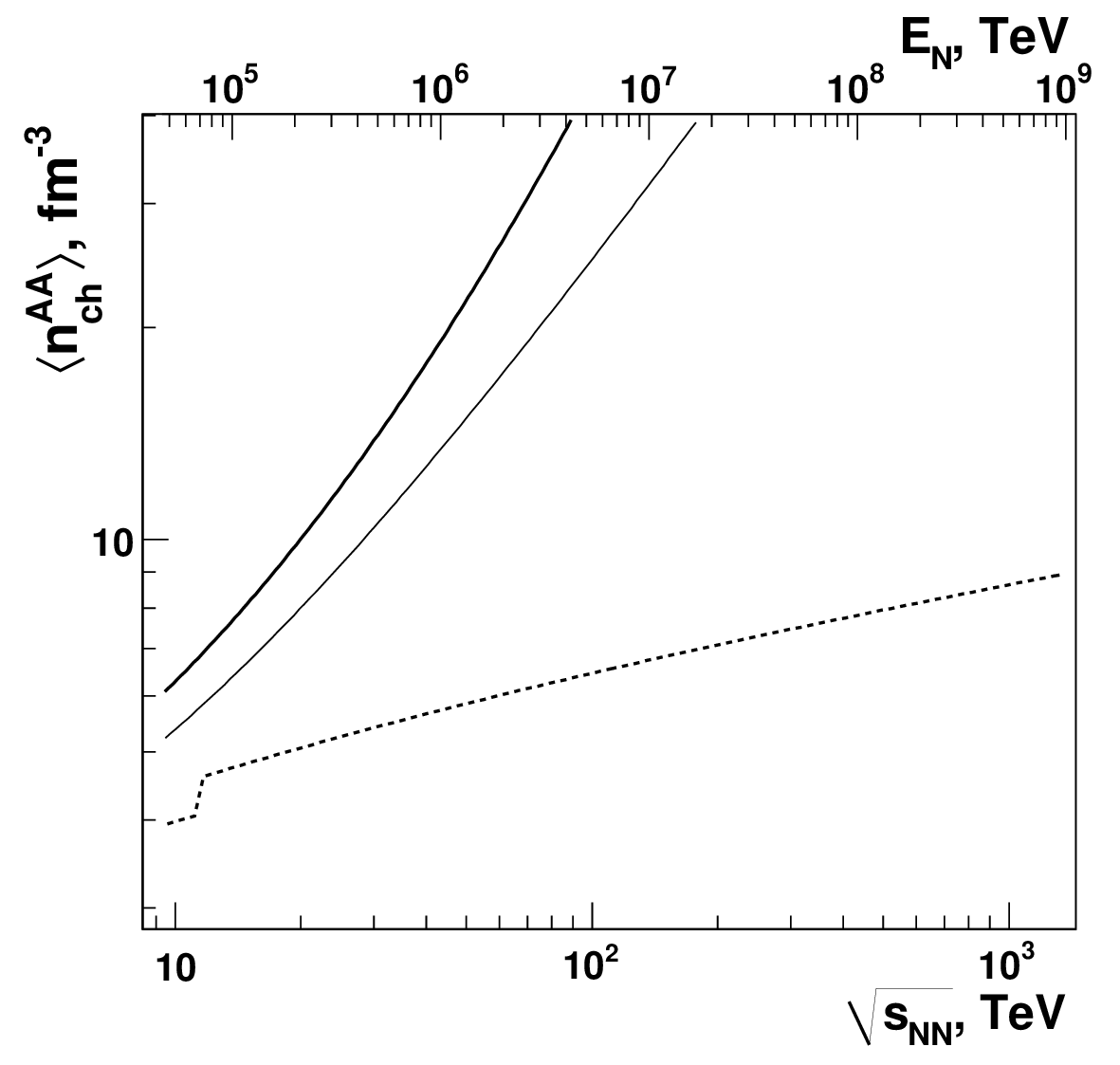}
\vspace*{8pt} \caption{Energy dependence of $\langle
n_{\scriptsize{\mbox{ch}}}\rangle$ in symmetric ($A+A$) heavy ion
collisions. The possible effect of the Bose--Einstein condensation
is taken into account in accordance with (\ref{eq:2.10}) with
$X=5$ for the energy region with $\langle
n_{\scriptsize{\mbox{ch}}}^{AA}\rangle
> n_{\scriptsize{\mbox{ch,c}}}^{AA}$. Solid line correspond to the
approximation (\ref{eq:2.6}) of $\langle
N_{\scriptsize{\mbox{ch}}}^{AA}\rangle$ with thin line shown for
the case of absence of BEC for completeness of the information.
Dashed line is for the approximation (\ref{eq:2.7}).}\label{fig:3}
\end{figure*}

Within the present work as the first stage of the quantitative
study of possible influence of BEC on the pion multiplicity at
ultra-high energies the simple approach of appropriate constant
$X$ is considered without taking into account the energy
dependence of the parameter due to definition in (\ref{eq:2.2}).
Fig. \ref{fig:3} demonstrates the energy dependence of $\langle
n_{\scriptsize{\mbox{ch}}}\rangle$ in symmetric ($A+A$) heavy ion
collisions with possible effect of BEC at appropriate condition,
i.e. for the energy region with $\langle
n_{\scriptsize{\mbox{ch}}}^{AA}\rangle
> n_{\scriptsize{\mbox{ch,c}}}^{AA}$. The influence of BEC is
taken into account in accordance with (\ref{eq:2.10}) and
corresponding curves are calculated at $X=5$. As seen in Fig.
\ref{fig:2} the condition $\langle
n_{\scriptsize{\mbox{ch}}}^{AA}\rangle
> n_{\scriptsize{\mbox{ch,c}}}^{AA}$ for the onset of BEC is valid
for the approximation (\ref{eq:2.6}) even at energies some smaller
than the low boundary $E_{N}=10^{17}$ eV of the considered range.
Therefore two curves, namely, with (solid line) and without (thin
solid line) accounting for possible influence of BEC are
calculated for the approximation (\ref{eq:2.6}) in order to clear
show the change of $\langle
n_{\scriptsize{\mbox{ch}}}^{AA}\rangle$ versus energy parameter
due to BEC. Dashed line is for the approximation (\ref{eq:2.7}) in
Fig. \ref{fig:3}. The BEC results in to the visible increase of
charged particle density at even large enough $X=5$ for the
appropriate energy range with $\langle
n_{\scriptsize{\mbox{ch}}}^{AA}\rangle
> n_{\scriptsize{\mbox{ch,c}}}^{AA}$. Moreover the increase due to
BEC amplifies with growth of the energy parameter for collision
process\footnote{The sharp irregular behavior of the dependence of
$\langle n_{\scriptsize{\mbox{ch}}}^{AA}\rangle$ versus energy, in
particular, for the approximation (\ref{eq:2.7}) in Fig.
\ref{fig:3} is mostly explained by the using of aforementioned
strict inequality for the exact median values of $\langle
n_{\scriptsize{\mbox{ch}}}^{AA}\rangle$ and
$n_{\scriptsize{\mbox{ch,c}}}^{AA}$. Accounting for the
uncertainties of the multiplicity parameters will result in the
creation of some finite energy range with $\langle
n_{\scriptsize{\mbox{ch}}}^{AA}\rangle \approx
n_{\scriptsize{\mbox{ch,c}}}^{AA}$ which will be considered as
some transition region. By analogy with BEC of ultra-cold atoms in
condense matter experiments \cite{RMP-74-875-2002} one can expect
the gradual amplification of the influence of BEC on $\langle
n_{\scriptsize{\mbox{ch}}}^{AA}\rangle$ with growth of the pion
multiplicity in this region. Therefore smoother behavior of energy
dependence of $\langle n_{\scriptsize{\mbox{ch}}}^{AA}\rangle$ is
expected, in general, in the finite energy range close to the
onset of BEC. Exact form of the curve $\langle
n_{\scriptsize{\mbox{ch}}}^{AA}\rangle(s_{NN})$ or vs $E_{N}$ in
energy domain with $\langle n_{\scriptsize{\mbox{ch}}}^{AA}\rangle
\approx n_{\scriptsize{\mbox{ch,c}}}^{AA}$ depends on dynamic
features of the creation of BEC in multiparticle production
processes and it is the subject of additional study.}.

Within the present work the quantitative characteristics are
determined
\begin{equation}
\displaystyle z_{\pi}^{(n)}=\frac{\ln \langle
n_{\scriptsize{\mbox{ch,BEC}}}^{AA}\rangle-\ln \langle
n_{\scriptsize{\mbox{ch,0}}}^{pp}\rangle}{\ln \langle
n_{\scriptsize{\mbox{ch,0}}}^{AA}\rangle-\ln \langle
n_{\scriptsize{\mbox{ch,0}}}^{pp}\rangle}, \label{eq:2.11}
\end{equation}
\begin{equation}
\displaystyle \Delta z_{\pi}^{(n)}=z_{\pi}^{(n)}-1.
\label{eq:2.12}
\end{equation}
Here $\langle n_{\scriptsize{\mbox{ch,BEC}}}^{AA / pp}\rangle$ is
the average density of charged particles (pions) with taking into
account the possible Bose--Einstein condensation effect at the
region of (kinematic) parameter space with average density larger
than critical one ($\langle n_{\scriptsize{\mbox{ch}}}\rangle >
n_{\scriptsize{\mbox{ch,c}}}$) in $A+A$ or $p+p$ collisions
respectively, $\langle n_{\scriptsize{\mbox{ch,0}}}^{AA /
pp}\rangle$ is the average particle density when the
Bose--Einstein effect is switched off in the fixed type
interaction. The parameters (\ref{eq:2.11}) and (\ref{eq:2.12})
are used here for quantitative study of the effect of
Bose--Einstein condensation on the density of secondary charged
pions and they are analogues of the corresponding parameters used
in the study of muon excess in the collisions of UHECR particles
with atmosphere \cite{EPJWC-210-02004-2019,PRD-107-094031-2023}.

\begin{figure*}
\includegraphics[width=16.0cm,height=16.0cm]{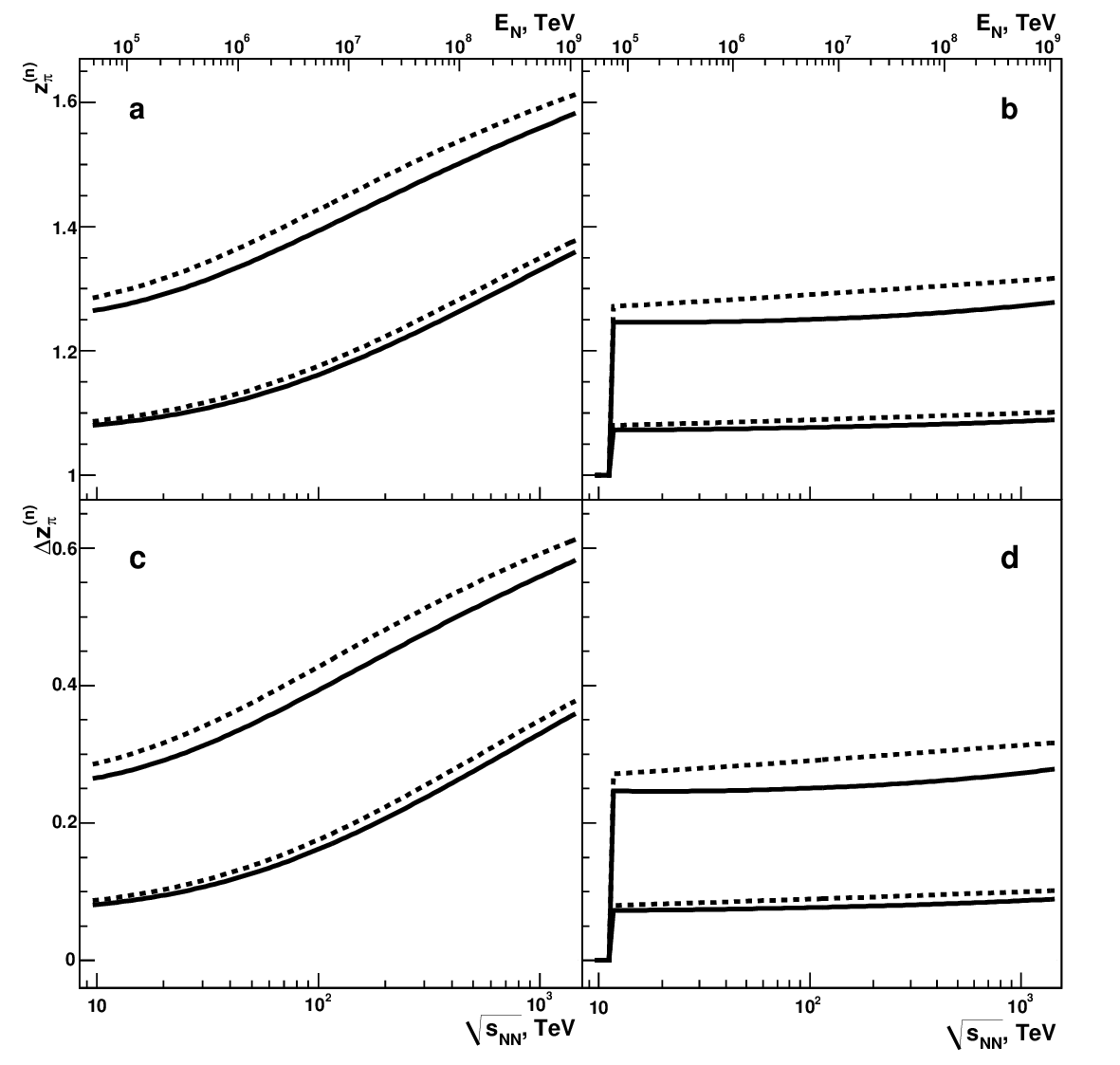}
\vspace*{8pt} \caption{Parameters $z_{\pi}^{(n)}$ (a, b) and
$\Delta z_{\pi}^{(n)}$ (c, d) estimated for charged pions with
help of $\langle n_{\scriptsize{\mbox{ch}}}\rangle$ for
appropriate ($p+p$, $A+A$) interactions in dependence on energy.
In the case of a symmetric ($A+A$) ion collisions the
approximation (\ref{eq:2.6}) is used for $\langle
N_{\scriptsize{\mbox{ch}}}^{AA}\rangle$ for the panels (a, c)
while analytic function (\ref{eq:2.7}) is used for the panels (b,
d). In each panel solid lines correspond to the equation
(\ref{eq:2.3}) for $\langle
N_{\scriptsize{\mbox{ch}}}^{pp}\rangle$, dashed lines are for
equation (\ref{eq:2.4}). Effect of BEC is taken into account in
accordance with (\ref{eq:2.10}) for energy region with $\langle
n_{\scriptsize{\mbox{ch}}}\rangle > n_{\scriptsize{\mbox{ch,c}}}$
in certain type of collisions if any. The upper collection of
curves corresponds to the $X=2$, lower curves are for
$X=5$.}\label{fig:4}
\end{figure*}

Fig. \ref{fig:4} shows $z_{\pi}^{(n)}$ (a, b) and $\Delta
z_{\pi}^{(n)}$ (c, d) in dependence on energy parameters. The
quantities (\ref{eq:2.11}) and (\ref{eq:2.12}) are calculated for
charged pions with help of corresponding
$n_{\scriptsize{\mbox{ch}}}$. In the case of a symmetric ($A+A$)
ion collisions the approximation (\ref{eq:2.6}) is used for
average total multiplicity $\langle
N_{\scriptsize{\mbox{ch}}}^{AA}\rangle$ for the panels (a, c)
while analytic function (\ref{eq:2.7}) is used for the panels (b,
d). In each panel solid lines correspond to the equation
(\ref{eq:2.3}) for $\langle
N_{\scriptsize{\mbox{ch}}}^{pp}\rangle$, dashed lines are for
equation (\ref{eq:2.4}). Effect of BEC is taken into account in
accordance with (\ref{eq:2.10}) for energy region with $\langle
n_{\scriptsize{\mbox{ch}}}^{AA}\rangle >
n_{\scriptsize{\mbox{ch,c}}}^{AA}$. The upper collection of curves
corresponds to the $X=2$, lower curves are for $X=5$. In general,
Figs. \ref{fig:4}a, b demonstrate that the curves for
$z_{\pi}^{(n)}$ versus energy show the close behavior for various
parameterizations of $\langle N_{\scriptsize{\mbox{ch}}}\rangle$
in $p+p$, especially at larger $X$. The functional behavior of
$z_{\pi}^{(n)}(s_{NN})$ is almost independent on value of $X$ for
any combinations of the approximations for $\langle
N_{\scriptsize{\mbox{ch}}}^{AA}\rangle$ and $\langle
N_{\scriptsize{\mbox{ch}}}^{pp}\rangle$ (Figs. \ref{fig:4}a, b).
The clear increase of $z_{\pi}^{(n)}$ is observed with growth of
energy in the case of the approximation (\ref{eq:2.6}) for
$\langle N_{\scriptsize{\mbox{ch}}}^{AA}\rangle$ (Fig.
\ref{fig:4}a) whereas there is almost no dependence
$z_{\pi}^{(n)}$ vs $s_{NN}$ $(E_{N})$ for the function
(\ref{eq:2.7}) especially at $X=5$ in the energy domain with the
presence of BEC effect (Figs. \ref{fig:4}b). Values of
$z_{\pi}^{(n)}$ are noticeably larger for calculations at $X=2$
with equation (\ref{eq:2.4}) for $\langle
N_{\scriptsize{\mbox{ch}}}^{pp}\rangle$ than that for the
approximation (\ref{eq:2.3}) in any considered cases of analytic
parameterization for $\langle
N_{\scriptsize{\mbox{ch}}}^{AA}\rangle$ vs energy. This
discrepancy is some clearer for the function (\ref{eq:2.7}) in the
domain $E_{N} \gtrsim 10^{19}$ eV (Fig. \ref{fig:4}b). As
expected, the features of the behavior of $\Delta z_{\pi}^{(n)}$
in dependence on energy parameters (Figs. \ref{fig:4}c, d) are the
same as well as the aforementioned observations for
$z_{\pi}^{(n)}$ at corresponding $X$ and choice of the
approximations for $\langle
N_{\scriptsize{\mbox{ch}}}^{AA/pp}\rangle$ due to relation
(\ref{eq:2.12}) between these parameters.

To obtain estimations of the parameters $z_{\pi}^{(n)}$
(\ref{eq:2.11}) and $\Delta z_{\pi}^{(n)}$ (\ref{eq:2.11}) based
on the measurements on accelerators with high statistical
precision the initial energies for the corresponding data samples
for $p+p$ and $A+A$ collisions should be equal or at least be
close to each other. Thus, based on the aforementioned, obtaining
of experimental results for the considered ultra-high energy range
$E_{p}=10^{17}-10^{21}$ eV can be expect only in a fairly distant
future, not earlier than mid-2060s at the conservation of the
planed schedule for implementation of the FCC--hh project
\cite{EPJST-228-755-2019}. That is extra emphasizes the importance
for study of the interactions of UHECR particles with nuclei for
the physics of fundamental interactions at energies unachievable
in accelerator physics in the near- and medium-term perspective.

It should be noted that $\Delta z_{\pi}^{(n)} > 0$ at any energies
under consideration (Figs. \ref{fig:4}c, d) and the present
quantitative analysis clear shows and supports the noticeable
increase of average density and, consequently, average total
multiplicity of charged pions in nuclear collisions at $E_{N} \geq
10^{17}$ eV due to influence of BEC. Of course, this statement is
only with taking into of median curves for the corresponding
densities in (\ref{eq:2.11}). Due to aforementioned decay modes of
$\pi^{\pm}$ the parameters (\ref{eq:2.11})and (\ref{eq:2.12}) can
be directly associated with the corresponding quantities for muon
yield as $z_{\pi}^{(n)} \approx z_{\mu}^{(n)}$ and $\Delta
z_{\pi}^{(n)} \approx \Delta z_{\mu}^{(n)}$. Then the range of
$\Delta z_{\mu}^{(n)}$ estimated within the present work with help
of $\Delta z_{\pi}^{(n)}$ (Figs. \ref{fig:4}c, d) agrees
reasonably on order of magnitude with the values of corresponding
parameter obtained within studies of the muon puzzle in UHECR
which are mostly in the range $\Delta z \sim 0.2-1.0$ with large
errors \cite{EPJWC-210-02004-2019,PRD-107-094031-2023}. This
agreement is for any choice of $\langle
N_{\scriptsize{\mbox{ch}}}^{AA/pp}\rangle$ considered here and for
energies $E_{N} \geq 10^{17}$ eV. The clear energy dependence of
$\Delta z$ is absent because of large uncertainties of
measurements \cite{EPJWC-210-02004-2019,PRD-107-094031-2023} with
some indication on the weak increase of $\Delta z$ with growth of
energy. The dependences of $\Delta z_{\pi}^{(n)}$ versus energy
shown in Fig. \ref{fig:4}c for $X=2$ and 5 qualitatively closer to
the general tendency in muon data
\cite{EPJWC-210-02004-2019,PRD-107-094031-2023} than other our
curves (Fig. \ref{fig:4}d). Thus the approximation (\ref{eq:2.6})
for $\langle N_{\scriptsize{\mbox{ch}}}^{AA}\rangle$ looks like
slightly more preferably than (\ref{eq:2.7}) for agreement with
available muon results for $\Delta z$ from
\cite{EPJWC-210-02004-2019,PRD-107-094031-2023}. But the large
errors of the muon data do not allow the exact exclusion of the
almost energy-independent behavior of $\Delta z$
\cite{EPJWC-210-02004-2019,PRD-107-094031-2023}. Therefore the
aforementioned statement is only qualitative and future
improvements are important for both the phenomenological results
obtained here and the precision of measurements of muon yields in
UHECR for more rigorous conclusion.

Based on the general properties of multiparticle production
process one can suggest at qualitative level that BEC will result
in the large number of soft pions and excess will be dominated
namely the soft pions. Consequently, the muons appearing via decay
of such pions will be soft too. This qualitative hypothesis means
that BEC effect could lead to muon yield with, in particular,
kinematic properties of the muons are rather different from those
in the case of decays  of heavy particles, in particular,
(anti)top quark \cite{Okorokov-JPCS-1690-012006-2020}. Moreover
the studies of the muon puzzle in UHECR mostly request the large
number of relatively soft muons and, at least, some of these works
consider soft sector of the strong interaction physics as the
possible field for the explanation of the muon puzzle, for
instance, various hadronization schemes
\cite{PRD-107-094031-2023}. All of these can be considered as, at
least, indirect and qualitative indication in favorable of BEC as
the possible source of muon excess at ultra-high energies rather
than decays of heavy particles but without full exclusion of the
last hypothesis. Therefore, BEC can be one of the sources for the
muon excess especially at highest energies and, consequently, one
of the possible solutions of the muon puzzle in UHECR.

There is wide space for the improvement of the phenomenological
model suggested within the present paper, in particular, regards
of the space-time extension of the emission region of secondary
pions, statistical properties of the final-state boson system
(better choice of type of distribution, of approach for
description of the Bose system) etc. This work is in the progress.
But it should be noted that the aforementioned qualitative
agreement is already achieved within the specific version of the
model considered here for the estimations of $\Delta
z_{\mu}^{(n)}$ discussed above and the results obtained for UHECR
\cite{EPJWC-210-02004-2019,PRD-107-094031-2023}. Therefore study
of possible BEC influence on the pion yield at ultra-high energies
can be considered as one of the promising ways for better
understanding of the source for the muon puzzle in UHECR.

\section{Conclusions}\label{sec:4}

Summarizing the foregoing, one can draw the following conclusions.

The phenomenological approach is suggested for quantitative study
the effect of Bose--Einstein condensation on the density of
secondary charged pions in general case of multiparticle
production process for wide initial energy range, particularly,
for ultra-high energy domain. The possible influence of BEC of
secondary pions within the model with complete $n$-particle
symmetrization has been studied for muon production in UHECR
particle collisions.

The results of this study indicate that hypothesis of BEC
corresponding to the lasing feature for pion production seems
unfavorable in $p+p$ collisions up to the ultra-high
center-of-mass energies $\mathcal{O}$(1 PeV). But symmetrization
effect would be affect on the charged particle multiplicity in
symmetric ($A+A$) heavy ion collisions at all energy range $E_{N}
\geq 10^{17}$ eV and phenomenological models for mean total
charged particle multiplicity under consideration. Both
statements, for $p+p$ and $A+A$, agree with our previous work
\cite{Okorokov-PAN-82-838-2019}.

Within the present work variables $z_{\pi}^{(n)}$ and $\Delta
z_{\pi}^{(n)}$ are determined in the similar manner to the
parameters used for the study of muon excess in the collisions of
UHECR particles. These variables allow the quantitative study of
difference between charged pion yields in the cases of presence of
the BEC and at absence of the effect under discussion. For
secondary pions the special case of the Poissonian multiplicity
distribution has been considered and corresponding
$z_{\pi}^{(n)}(s)$, $\Delta z_{\pi}^{(n)}(s)$ dependence has been
derived for various relations between analytic approximations of
average total multiplicities in $p+p$ and $A+A$ interactions. The
behavior of energy dependence of $z_{\pi}^{(n)}$ and $\Delta
z_{\pi}^{(n)}$ varies with approximation of average total
multiplicity in $A+A$ collisions. The both parameters show the
increase of pion yield for the case of presence of Bose--Einstein
condensation and magnitude of this increase does not contradict,
at least, at qualitative level to the muon excess observed in the
collisions of UHECR particles.

Therefore the novel feature of multiparticle production processes
-- Bose--Einstein condensation -- could be, in general, contribute
to the muon yield measured in the collisions of ultra-high energy
cosmic ray particles with atmosphere.

\section*{Acknowledgments}\label{sec:4}

This work was supported in part within the National Research
Nuclear University MEPhI Program ''Priority 2030".

\newpage
\newpage
\newpage
\newpage

\begin{thebibliography}{99}
\bibitem{Okorokov-PAN-82-838-2019}
V.~A.~Okorokov, Phys. At. Nucl. \textbf{82}, 838 (2019).
\bibitem{NIMA-742-228-2014}
A.~A.~Petrukhin, Nucl. Instr. Meth. Phys. Res. A \textbf{742}, 228
(2014).
\bibitem{EPJWC-210-02004-2019}
H.~P.~Dembinski et al., Eur. Phys. J. Web Conf. \textbf{210},
02004 (2019).
\bibitem{PRD-107-094031-2023}
S.~Baur et al., Phys. Rev. D \textbf{107}, 094031 (2023).
\bibitem{arXiv-2205.05845-astro-ph.HE-2022}
A.~Coleman et al., Astropart. Phys. \textbf{149}, 102819 (2023). 
\bibitem{book-SM-1-1987}
L.~D.~Landau and E.~M.~Lifshitz, \emph{Course of theoretical physics}
\textbf{V}, \emph{Statistical physics. Part 1.} (Oxford, Pergamon Press,
1980), p. 181; \emph{ibid} \textbf{IX}, \emph{Statistical
physics. Part 2.} (Oxford, Pergamon Press,
1980), p. 99; K.~Huang,
\emph{Statistical mechanics} (N.J., J.Wiley and Sons, 1987), p.
287.
\bibitem{FE-1-219-1988}
D.~P.~Zubarev, Bose--Einstein condensation. \emph{Physical
encyclopiedia} \textbf{I}, (M., Large Russian Encyclopiedia, 1988),
p. 219.
\bibitem{Universe-9-411-2023}
D.~Anchishkin, V.~Gnatovskyy, D.~Zhuravel, V.~Karpenko,
I.~Mishustin and H.~Stoecker, Universe \textbf{9}, 411 (2023).
\bibitem{RMP-71-463-1999}
F.~Dalfovo, S.~Giorgini, L.~P.~Pitaevskii and S.~Stringari, Rev.
Mod. Phys. \textbf{71}, 463 (1999); Y.~Castin, arXiv:
cond-mat/0105058, 2005.
\bibitem{PRA-54-656-1996}
W.~Ketterle and N.~J.~van Druten, Phys. Rev. A\textbf{54}, 656
(1996).
\bibitem{PLB-301-159-1993}
S.~Pratt, Phys. Lett. B \textbf{301}, 159 (1993).
\bibitem{PRL-80-916-1998}
T.~Cs{\"{o}}rg{\'{o}} and J.~Zim{\'{a}}nyi, Phys. Rev. Lett.
\textbf{80}, 916 (1998).
\bibitem{HIP-9-241-1999}
J.~Zim{\'{a}}nyi and T.~Cs{\"{o}}rg{\'{o}}, Heavy Ion Phys.
\textbf{9}, 241 (1999).
\bibitem{PTEP-2022-083C01-2022}
R.~L.~Workman et al. (Particle Data Group), Prog. Theor. Exp.
Phys. \textbf{2022}, 083C01 (2022).
\bibitem{Okorokov-PAN-81-508-2018}
V.~A.~Okorokov, Yad. Fiz. \textbf{81}, 481 (2018) [Phys. At. Nucl.
\textbf{81}, 508 (2018)].
\bibitem{Greisen-PRL-16-748-1966}
K.~Greisen, Phys. Rev. Lett. \textbf{16}, 748 (1966);
G.~T.~Zatsepin and V.~A.~Kuzmin, JETP Lett. \textbf{4}, 78 (1966).
\bibitem{Okorokov-AHEP-2016-5972709-2016}
V.~A.~Okorokov, Adv. High Energy Phys. \textbf{2016}, 5972709
(2016).
\bibitem{PRC-73-034905-2006}
J.~Cleymans, H.~Oeschler, K.~Redlich and S.~Wheaton, Phys. Rev. C \textbf{73}, 034905 (2006).
\bibitem{PRD-93-054046-2016}
E.~K.~G.~Sarkisyan, A.~N.~Mishra, R.~Sahoo and A.~S.~Sakharov,
Phys. Rev. D \textbf{93}, 054046 (2016).
\bibitem{JPGNNP-37-083001-2010}
J.~F.~Grosse-Oetringhaus and K.~Reygers, J. Phys. G: Nucl. Part.
Phys. \textbf{37}, 083001 (2010).
\bibitem{PLB-459-341-1999}
I.~M.~Dremin and J.~W.~Gary, Phys. Lett. B \textbf{459}, 341
(1999).
\bibitem{PR-349-301-2001}
I.~M.~Dremin and J.~W.~Gary. Phys. Rev. \textbf{349}, 301 (2001).
\bibitem{EPJC-35-457-2004}
A.~Heister et al. (ALEPH Collab.), Eur. Phys. J. C \textbf{35},
457 (2004).
\bibitem{PLB-726-610-2013}
E.~Abbas et al. (ALICE Collab.) Phys. Lett. B \textbf{726}, 610
(2013).
\bibitem{Okorokov-AHEP-2015-790646-2015}
V.~A.~Okorokov, Adv. High Energy Phys. \textbf{2015}, 790646
(2015).
\bibitem{EPJST-228-1109-2019}
A.~Abada et al. (FCC Collab.), Eur. Phys. J. Spec. Topics
\textbf{228}, 1109 (2019).
\bibitem{EPJST-228-755-2019}
A.~Abada et al. (FCC Collab.), Eur. Phys. J. Spec. Topics
\textbf{228}, 755 (2019).
\bibitem{PPNP-63-293-2009}
J.~Bl$\ddot{\mbox{u}}$mer, R.~Engel, and
J.~R.~H$\ddot{\mbox{o}}$randel, Prog. Part. Nucl. Phys.
\textbf{63}, 293 (2009).
\bibitem{PRD-103-103009-2021}
P.~Lipari, Phys. Rev. D \textbf{103}, 103009 (2021); N.~Arsene,
Universe \textbf{7}, 321 (2021).
\bibitem{PRD-90-012012-2014}
A.~Aab et al. (Pierre Auger Collab.), Phys. Rev. D \textbf{90},
012012 (2014). Addendum: Phys. Rev. D \textbf{90}, 039904 (2014).
Erratum: Phys. Rev. D \textbf{92}, 019903 (2015).
\bibitem{RMP-74-875-2002}
E.~A.~Cornell and C.~E.~Wieman, Rev. Mod. Phys. \textbf{74}, 875
(2002); W.~Ketterle, \emph{ibid} \textbf{74}, 1131 (2002);
E.~Streed ey al., Rev. Sci. Instrum. \textbf{77}, 023106 (2006).
\bibitem{PRC-93-054908-2016}
J.~Adam et al. (ALICE Collab.), Phys. Rev. C \textbf{93}, 054908
(2016).
\bibitem{Okorokov-JPCS-1690-012006-2020}
V.~A.~Okorokov, J. Phys. Conf. Ser. \textbf{1690}, 012006 (2020);
Phys. At. Nucl. \textbf{86}, 742 (2023).

\end{thebibliography}
\end{document}